\documentclass[10pt,letterpaper,twocolumn]{article} 

\usepackage{ol2} 

\newcommand{\eeqref}[1]{Eq.~(\ref{#1})}

\newcommand{\slsh}
{
\hspace{0.5pt} / \hspace{-0.5pt}
}

\begin{document}

\twocolumn[

\title{Measurement of superluminal phase and group velocities of Bessel beams in free space}

\author{B. Braverman,$^{2}$ K. B. Kuntz,$^{1}$ M. Lobino,$^{1}$ E. M. Pessina,$^3$ and A.I. Lvovsky$^{1,*}$}

\address{$^1$Institute for Quantum Information Science, University of Calgary, Calgary AB T2N 1N4 Canada}
\address{$^2$Department of Physics, University of Toronto, Toronto ON M5S 1A7 Canada}
\address{$^3$Pirelli Labs, Viale Sarca 336, 20126, Milano, Italy}
\address{$^*$Corresponding author: lvov@ucalgary.ca}

\maketitle

\begin{abstract} We report on an interferometry-based measurement of
the phase and group velocities of optical Bessel beams, providing
confirmation of their superluminal character in the non-diffractive
region. The measurements were performed in free space with a
continuous wave laser and femtosecond pulses for phase and group
velocities respectively. The Bessel beams were produced using a conical mirror. \end{abstract}


]

Einstein's special theory of relativity prohibits communicating at a
speed faster than light. Nevertheless, superluminal phase and group
velocities of electromagnetic waves have been observed
in a variety of settings \cite{Basov,ChuWong,Steinberg,Kuzmich,Boyd}. While, for different reasons, the
observed behavior of optical waves cannot be used for superluminal
signalling, these studies captured significant attention both within
and outside the optics community and lead to a more precise
understanding of the notion of signal velocity.

Among the many arrangements where light exhibits quasi-superluminal
properties, a special place belongs to laser beams whose radial
profile is given by the zeroth-order Bessel function (Bessel
beams). Discovered by Durnin in 1987 \cite{Durnin:87}, the Bessel beam [Fig.~1(a)] is a
coherent superposition of an equal-phase set of infinite plane waves
that propagate at the cone angle $\theta$ relative to a fixed axis
(which we define as $z$). The component waves interfere to produce a wavefront which is invariant in the propagation direction, and is thus ``diffraction-free'', albeit over a limited propagation range of $R \slsh \tan{\theta}$, where $R$ is the radius of the optical element used to generate the beam. In the first realization, Bessel beams
were produced using a circular slit and a lens \cite{DurninME:87},
but subsequently a refracting axicon \cite{Indebetouw:89} became the
most common approach, and recently a conical mirror scheme has been
proposed \cite{Fortin:03}.

While each component mode of the Bessel wave propagates at speed $c$,
the wavevector of the Bessel wavefront is equal to the $z$-projection of each
component's wavevector, i.e. $k=(\omega/c)\cos\theta$, where
$\omega$ is the optical frequency. Accordingly, the phase and group
velocities of the Bessel wavefront are $v=\omega/k=c/\cos\theta$,
exceeding the speed of light in vacuum. In contrast to Refs.~\cite{Basov,ChuWong,Steinberg,Kuzmich,Boyd}, the superluminal property of Bessel beams is not related to absorption/dispersion anomalies in the spectrum of the propagation medium but arises due to unusual interference between the plane wave components of the beam. As a result, superluminal propagation is observed in a wide spectral range.

As in the case of dispersive media, superluminal velocity of the light wave does not imply that faster-than-light signaling is possible \cite{Lunardi:01,MugnaiComment1,MugnaiComment2}. Indeed, the
electromagnetic fields at points $A$ and $B$ on the Bessel beam
axis [Fig.~1(a)] form due to interference of conical waves emanating
from different points on the generating optical element ($A',A''$
and $B',B''$, respectively). There is thus no causal connection
between the signal arriving at points $A$ and $B$, so the
superluminal behavior does not violate causality and is not in
contradiction with special relativity. We also note that the superluminal character of the Bessel beam is limited to the same propagation range of $R \slsh \tan{\theta}$ as is its non-diffracting profile.

Superluminal group velocity of the Bessel beam has been previously
reported \cite{MugnaiRR:00,LloydM:02,AlexeevKM:02}. However, none of
these papers conclusively confirm the expected superluminality in
free space, in part due to large measurement uncertainties
\cite{MugnaiRR:00,LloydM:02} or the usage of an ionizable medium
\cite{AlexeevKM:02}. In addition, Ref.~\cite{LloydM:02} reported superluminal phase velocity; however, the experimental results differed greatly from theoretical predictions. In this Letter, we report, for the first time, a measurement of the phase and group velocities of Bessel beams in free space using a new interferometric
technique, confirming the expected value $c/\cos{\theta}$ for the
phase and group velocities with a high precision, and demonstrating
their superluminal character unambiguously.

\begin{figure}[tbp]
\centering
\includegraphics[width=0.9\columnwidth]{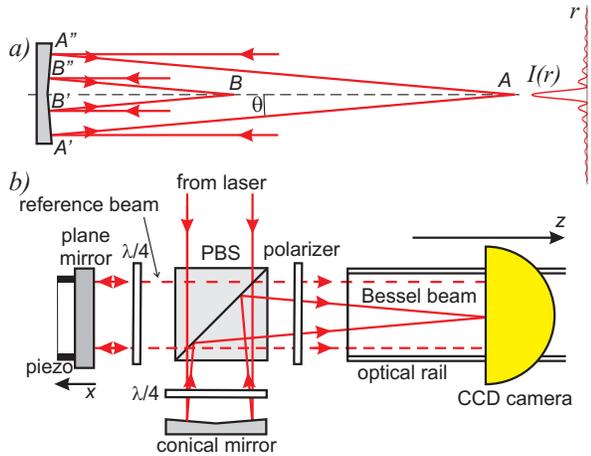}
\caption{a) Formation of the Bessel beam from a plane wave by a conical mirror and a plot of its radial intensity profile. b) Schematic of experimental apparatus used for phase and group velocity measurements. PBS, polarizing beam splitter.}
\label{fig:apparatus}
\end{figure}

We produced the Bessel beam using a silver coated conical mirror, which allowed us
to avoid dispersive distortion of ultrashort pulses. The mirror
had a diameter of 25.4 mm and an apex
angle of $\pi-\theta=179^\circ$. The mirror substrate was manufactured
by diamond turning technique. The key component of
the experimental apparatus [Fig. \ref{fig:apparatus}(b)] was a Michelson interferometer,
with the conical mirror in one arm and a plane mirror in the other,
allowing us to compare the propagation of ordinary Gaussian
wavefronts with those of a Bessel beam \cite{SaariR:97}.

In order to measure the phase velocity of the Bessel beam, we used the
radiation of a continuous-wave He:Ne laser operating at 543.5 nm.
The laser output beam was expanded by two telescopes, 25x total, to
produce a collimated beam 35 mm in diameter, wider than the conical mirror, and entered the Michelson interferometer. The
two quarter-wave plates were rotated to achieve similar intensities of
the reference beam and the central peak of the Bessel beam, resulting
in an easily visible interference pattern, which was detected by a
CCD camera mounted on an optical rail.

The reference and Bessel waves have equal frequency but slightly
different phase velocities along the $z$ axis, respectively $v_{pR}$ and $v_{pB}$. They will
produce interference fringes of period $\Delta z = {2\pi v_{pB} v_{pR}} / (\omega |\Delta v_{p}|)$, where $\Delta v_{p}=v_{pR}-v_{pB}$. If $|\Delta v_{p}|
\ll v_{pR} \approx c$, then
\begin{equation}\label{eq:phaseequation}
{\Delta
z}=\frac{\lambda c}{|\Delta v_{p}|}.
\end{equation}
However, the observation of interference fringes can yield only the absolute value of
$\Delta v_{p}$, giving no information about its sign.

We solve this problem by varying the angle $\phi$ between the wave
vectors of the reference and Bessel beams, thus changing the $z$
component of the reference beam's phase
velocity according to $v_{pR}=c\slsh\cos{\phi}$. Observing the fringe spacing as a function of $\phi$, we can now determine the sign of $\Delta
v_{p}$. Specifically, $\Delta z$ will increase with increasing $\phi$ for small $\phi$ if and only if the Bessel beam's phase velocity exceeds that of the reference beam.

The angle $\phi$ was manipulated by aligning the reference beam parallel to the optical rail and turning the conical mirror slightly. The relative propagation direction of the beams, and hence the value of $\phi$, was determined by measuring the transverse positions of the centers of the two beams at different locations of the CCD on the optical rail. The fringe spacing $\Delta z$ was measured by finding the positions of interference minima between the reference beam and the central peak of the Bessel beam over several tens of interference periods and applying a linear fit.

\begin{figure}[tbp]
\centering
\includegraphics[width=0.9\columnwidth]{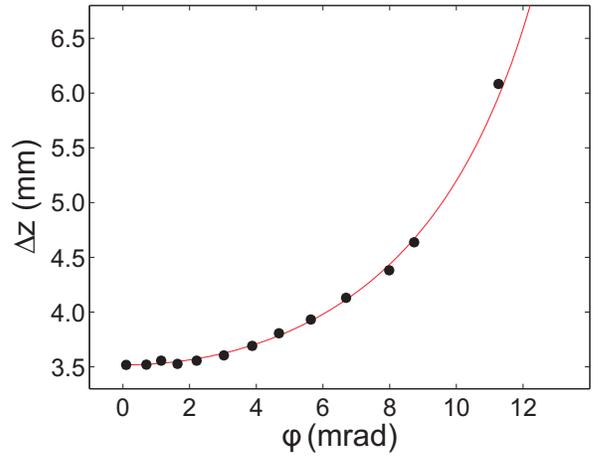}
\caption{Interference fringe period ($\Delta z$) as a function of the beam misalignment ($\phi$), and theoretical fit using Eq. (\ref{eq:phaseequation}) with $\theta=1.009^{\circ}$ as the only fitting parameter. }\label{fig:phaseresults}
\end{figure}

The results of this experiment are shown in Fig.~\ref{fig:phaseresults}. The fringe spacing increases with $\phi$, indicating that $v_{pB}>c$. A fit to the experimental points with \eeqref{eq:phaseequation} yields $v_{pB}=\left(1.000155\pm0.000003\right)c$, corresponding to a cone angle of $\theta=1.009^\circ\pm0.01^\circ$.


In order to measure the group velocity of the Bessel beam, the setup was slightly modified, as follows.  The He:Ne laser was replaced with a mode-locked Ti:Sapphire laser producing infrared pulses with $\lambda= 777$ nm, and 700 fs duration, and the optics were replaced accordingly to work with a different wavelength. In addition, the wider beam produced by the Ti:sapphire laser required only one telescopic expansion stage, which produced a beam 40mm in diameter. The plane mirror of the Michelson interferometer was mounted on a piezoelectric transducer and a micrometer translation stage, and the reference and Bessel beams were aligned to better than 0.5 mrad.

The slight difference in the group velocities results in the Bessel beam pulse gradually overtaking the reference pulse as they co-propagate. At one location along the $z$ axis, the two pulses are maximally temporally overlapped. Conversely, for each position of the camera along the $z$ axis, a specific relative delay between the two pulses, controlled by the position $x$ of the plane mirror, will produce maximum temporal overlap on the camera's photodiode array.

\begin{figure}[tbp]
\centering
\includegraphics[width=0.9\columnwidth]{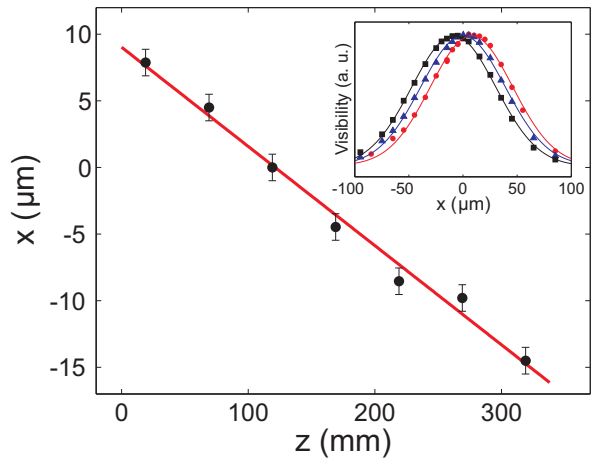}
\caption{Plane mirror position ($x$) required for maximizing the interference visibility as a function of the CCD position ($z$): ($\bullet$) experimental points and (line) linear fit. Inset: samples of normalized visibility data used to calculate the values $x$ for the dots in the main figure at $z=219$, 119 and 19 mm (left to right).}
\label{fig:groupresults}
\end{figure}

To measure the degree of temporal overlap of the pulses when detected by the CCD, we employed linear interferometric autocorrelation. If the pulses are temporally distinct at the CCD, the intensity profile observed will be the sum of the intensity profiles of the two beams. However, if the pulses arrive simultaneously, they produce an interference pattern that is sensitive to the relative phase between the beams. This allowed us to quantify the degree of temporal overlap of the pulses by observing the interference visibility at the center of the Bessel beam when the plane mirror was scanned back and forth over a distance of 1--1.5 wavelengths by the piezoelectric transducer. Sample data of these measurements can be seen in the inset to Fig.~\ref{fig:groupresults}.

For each position $z$ of the CCD, we adjusted the translation stage and determined the position $x$ of the middle of the scanning range where the visibility is maximized. This information allowed us to determine the group velocity of the Bessel pulse. Suppose that the CCD is translated away from the PBS by distance $\Delta z$, and to maintain maximum visibility, the delay of the reference pulse must be adjusted by lengthening the interferometer arm with the plane mirror by a distance $\Delta x$. This means that in the time that the Bessel pulse travels an extra distance of $\Delta z$ at velocity $v_{gB}$, the reference pulse travels $\Delta z + 2 \Delta x$ at speed $c$, giving
\begin{equation}
 \frac{\Delta z}{v_{gB}} = \frac{\Delta z + 2 \Delta x}{c}.
\label{eq:groupequation}
\end{equation}
We use this equation to determine $v_{gB}$.

As can be seen from Fig. \ref{fig:groupresults}, the relation between $x$ and $z$ is indeed linear, with the slope $\frac{\partial x}{\partial z} = (-7.5\pm0.3)\times 10^{-5}$. It is important to note that with increasing overall travel distance (i.e. larger $z$ values), the relative path length of the reference beam had to be reduced (i.e. decreasing $x$) to maintain maximum interference visibility at the position of the camera, implying that the Bessel beam was propagating at a greater velocity. Hence $v_{gB} = (1-2\frac{\partial x}{\partial z})c=\left(1.000150\pm0.000006\right)c$. This value is clearly superluminal, and corresponds to a cone angle of $\theta=0.992^\circ\pm0.02^\circ$.

In conclusion, we have for the first time measured the superluminal phase and group velocities of a Bessel beam in free space using a new interferometric approach that allowed us to quantify small velocity differences, both for continuous-wave and pulsed light. We have demonstrated the superluminal properties of Bessel beams in free space unambiguously, finding $v_{pB}=\left(1.000155\pm0.000003\right)c$ and $v_{gB}=\left(1.000150\pm0.000006\right)c$, confirming the value $c/\cos{\theta}=1.000152c$ predicted by theory for both phase and group velocities within narrow bounds.

\end{document}